\documentclass[10pt,letterpaper]{article}
\usepackage{hyperref}
\usepackage{amsmath}
\usepackage{subfigure}
\usepackage{epsfig}

\begin{document}

\title{\bf Compact-sized and broadband carpet cloak and free-space cloak}

\author{Hui Feng Ma, Wei Xiang Jiang, Xin Mi Yang, Xiao Yang Zhou, and Tie Jun Cui\footnote{tjcui@seu.edu.cn}\\
{\small\it State Key Laboratory of Millimeter Waves and Institute of Target Characteristics and Identification}\\
{\small\it Department of Radio Engineering, Southeast University,
Nanjing 210096, P. R. China.}}

\date{}
\maketitle

\begin{abstract}

Recently, invisible cloaks have attracted much attention due to
their exciting property of invisibility, which are based on a solid
theory of transformation optics and quasi-conformal mapping. Two
kinds of cloaks have been proposed: free-space cloaks, which can
render objects in free space invisible to incident radiation, and
carpet cloaks (or ground cloaks), which can hide objects under the
conducting ground. The first free-space and carpet cloaks were
realized in the microwave frequencies using metamaterials. The
free-space cloak was composed of resonant metamaterials, and hence
had restriction of narrow bandwidth and high loss; the carpet cloak
was made of non-resonant metamaterials, which have broad bandwidth
and low loss. However, the carpet cloak has a severe restriction of
large size compared to the cloaked object. The above restrictions
become the bottlenecks to the real applications of free-space and
carpet cloaks. Here we report the first experimental demonstration
of broadband and low-loss directive free-space cloak and
compact-sized carpet cloak based on a recent theoretical study. Both
cloaks are realized using non-resonant metamaterials in the
microwave frequency, and good invisibility properties have been
observed in experiments. This approach represents a major step
towards the real applications of invisibility cloaks.

\hskip 1.0mm


\end{abstract}

\newpage

Recently, a new theory of transformation optics and quasiconformal
mapping has been proposed to design the electromagnetic
metamaterials to control the paths of electromagnetic waves [1,2,4].
Based on the physical principle, an electromagnetic wave will always
travel in the quickest route between two points. In a homogeneous
and isotropic material, the route is a straight line. In an
inhomogeneous material, the route becomes nonlinear to make the
total traveling time be minimal because the wave travels at
different speeds inside. Hence one can control the route of wave by
designing the material parameters, which has found a lot of
potential applications such as cloaking objects invisible [1-7],
concentrating of electromagnetic waves [8,9], rotating of
electromagnetic waves [10], bending of electromagnetic waves [11],
and forming multi-beam antennas [12]. Artificial metamaterials
provide flexible choices of the designed parameters.

Among above potential applications, the invisibility cloak is the
most attractive. There have been two kinds of invisibility cloaks
proposed: free-space cloaks and carpet cloaks. A free-space cloak is
a designed material shell covering an object in free space, which
can guide the waves to propagate around the shell, making the object
inside invisible [1]. The theoretical tool to study invisible cloak
is the so-called transformation optics [13], which is based on the
coordinate transformation [14]. An optical conformal mapping method
was used to design the material parameters that create perfect
invisibility [1,2], in which the permittivity tensor $\epsilon$ and
the permeability tensor $\mu$ are both spatially varying and
anisotropic with singular values. By implementing such complicated
full material properties, the concealed object and the cloak appear
to have the free-space properties exactly when viewed externally,
which has been verified by full-wave simulations [15] and analytic
solutions [16].

However, the fully inhomogeneous and anisotropic parameters for
perfect free-space cloaks require extremely complicated metamaterial
design. For a two-dimensional (2D) cylindrical cloak in free space,
when the electric field is polarized along the cylinder axis, the
full parameters can be greatly simplified [3]. Since the reduced
constitutive parameters provide the same dispersion equation as the
full parameters for the 2D cloak, the electromagnetic waves have the
same trajectory inside the cloak with a penalty of nonzero
reflections. Using the reduced parameters, the first practical
free-space cloak has been realized in the microwave frequency [3].
In the experiment, a conducting cylinder was hidden inside the
cloak, which was constructed with the use of artificially structured
metamaterials. Due to the requirement of very large and very small
values of refraction index, resonant structures [17] have been
applied. This makes the invisible cloak operate in a narrow
frequency band with a relatively large loss. The experiment and
simulation results have a good match. But notable reflections and
shadow of the hidden object were observed due to the use of
simplified parameters and relatively large loss [3]. The narrow
bandwidth and large loss become the big restriction to real
applications of free-space cloaks.

In view of the difficulty to realize the free-space cloaks, a
recently published theory has suggested a carpet cloak, or
ground-plane cloak, which can hide any objects under a metamaterial
carpet [4]. Different from the free-space invisibility cloak, which
crushes the cloaked object to a point, the carpet cloak crushes the
hidden object to a conducting sheet [4]. In the other word, any
objects hidden under the carpet appear as a flat conducting sheet,
and hence cannot be detected by any exterior sources. The great
advantage of the carpet cloak is that it does not require singular
values for the material parameters [1]. By choosing appropriate
coordinate transformation, it has been shown that the anisotropy of
the cloak can be minimized and the range of the permittivity and
permeability is much smaller than that for the free-space cloak [4].
Hence the carpet cloak is easier to be realized using metamaterials.

The carpet cloak has very important applications, especially in the
microwave frequencies. For example, it can be used to hide the
aircraft on the airport, and the automobile on the ground, etc. The
carpet cloaks have been experimentally demonstrated firstly in the
microwave frequency [5], and then in the optical frequency [6,18],
using non-resonant metamaterials. The non-resonant metamaterials
have very low loss and operate in a broad frequency band, hence the
carpet cloaks have good invisibility performance [5,6]. However,
there are two restrictions in the existing carpet-cloak experiments:
1) the background is not free space to get easily realized material
parameters (in the microwave experiment, the refractive index of
background material was chosen as 1.331 [5]; in the optical
experiment, the refractive index of background material was chosen
as 1.58 [6]); 2) the carpet cloak has a very large size (in the
microwave experiment, the area ratio of the carpet cloak to the
cloaked region is as high as 65.7 [5]). The above restrictions make
the carpet cloak be difficult in real applications.

Here, we demonstrate experimentally the first broadband and low-loss
directive free-space cloak and compact-sized carpet cloak in the
free-space background. The experiments are based on a recent
theoretical study [7], which proposed a simplified carpet cloak (or
ground-plane cloak) made of only a few blocks of all-dielectric
isotropic materials. In our experiments, the simplified carpet cloak
is realized using non-resonant metamaterials in the microwave
frequency, which has a much smaller size than the full carpet cloak
reported in Ref. [5]. The area ratio of the cloak to the cloaked
region is as small as 7.33, which is only 11.2\% of the full cloak.
However, the presented carpet cloak has equally good invisibility
performance. Furthermore, a broadband and low-loss directive
free-space cloak is realized, that cloaks the radiation originating
from specified directions. Hence the presented work makes a major
step towards the real applications of invisibility cloaks.

For the design of the simplified carpet cloak, we consider a 2D
problem in which all the fields are invariant in the $z$ direction
and the electric fields are $z$-polarized. The cloaked object is a
car model, which is placed under a triangular conducting region. The
reason to choose the triangular shape is for easy installation in
real applications. We want to design a compact-sized carpet cloak
which is covering on the triangular region to make the whole system
appear as the original flat conducting plane as if the car does not
exist. We choose the shape of cloak as a rectangle with the width
125 mm and hight 50 mm except that the bottom boundary is partially
triangularly-shaped to place the car. Hence the car model is
concealed between the carpet cloak and the conducting plane as shown
in Fig. 1(c).

To design a compact-sized carpet cloak, the process of parameter
calculation contains two stages. First, we followed the technique
presented in [4] to get the complete cloaking region, in which a
quasi-conformal coordinate mapping is generated by minimizing the
Modified-Liao functional [20, 21] upon slipping boundary condition.
The refractive-index distribution of the cloak is computed
numerically, as shown in Fig. 1(a), in which the maximum anisotropic
factor is 1.07. Secondly, we use the grid optimization by recursive
division of Cartesian cells to sample the original refractive-index
distribution. In such a course, the small regions near the base
corners of the triangular object where the refractive-index is
smaller than 1 are treated as free space. Since the background of
the designed cloak is free space, the transformation generates cells
in region I (Fig. 1(a)) whose refractive indices are nearly 1. Hence
such a region can be simplified as free space, and the cloak will
reduce to the region II (Fig. 1(a)). The refractive index
distribution of the final carpet cloak is illustrated in Fig. 1(b),
in which the refractive indices are varied from 1 to 1.69.

The compact-sized carpet cloak is divided to 3-by-3-mm squares with
non-resonant elements shown in Figs. 1(c) and 1(d). By changing the
dimension $h$ (see Fig. 1(d)) from 0 to 2.2 mm, we are able to span
the required index range from $n =$ 1.07 to 1.87 at 10 GHz. After a
well-established retrieval process, the effective permittivity and
permeability for a given element can be found by numerical
simulation. Fig. 1(d) shows the curve that relates the refractive
index to a given element with the length $h$. Since the cell-to-cell
change in dimension is minor, the impedance is matched gradually in
the whole cloak over the entire measurement frequency range. The
compact-sized carpet cloak is fabricated on copper-clad printed
circuit board (PCB) with F4B substrate (the substrate thickness is
0.25 mm, with a dielectric constant of 2.65 and loss tangent of
0.001). The carpet has a size of $125\times 50$ mm$^{2}$ with a
height of 12 mm. The cloaking transformation is specifically
designed to compensate the triangular concealed region with the
height 13 mm and the bottom 125 mm.

To validate the invisible effect of the simplified carpet cloak , we
make use of a near-field scanning system to map the electric-field
distribution within a planar waveguide. In the planar waveguide, the
wave polarization is restricted to transverse electric. The
electric-field maps of the scattering region, including the
collimated incident and scattered beams, are shown in Fig. 2. The
incident waves are launched into the chamber from a standard X-band
coax-to-waveguide coupler to produce a nearly collimated microwave
plane beam. The beam is arbitrarily chosen to be incident on the
ground plane at an angle of 45$^{\circ}$ with respect to the normal.

Figure 2 shows the measured near-field distributions and the
far-field patterns of the carpet cloak. From Figs. 2(a) and (b),
there is considerable difference between the reflections from the
flat and curved surfaces. The beam reflected from the uncloaked
triangular bump shows the strong scattering of the bump. The flat
surface illustrates the expected collimation beam profile, similar
to that of the incident wave. To hide the bump on the surface, the
designed carpet cloak was placed around the bump and the measured
result is shown in Fig. 2(c). Subsequently, similar to the flat
reflecting surface, a single reflected beam is observed. This
demonstrates that the cloak has successfully transformed the curved
surface into a flat surface, giving the observer the impression that
the beam was reflected from a flat surface. Owing to the fact that
there is no penetration of light into the bump, the car model could
be placed behind it and effectively hidden, making the car
invisible. The far-field patterns are plotted in Figs. 2(g)-(i). For
the ground plane and the cloaked triangular concealed region, the
beam profiles show a very similar single peak; that is, the good
cloaking performance is observed. Unlike the above cases, the bump
alone exhibits a multi-peak far-field pattern (Fig. 2h), indicating
the strong perturbation of the beam.

To verify the broadband properties of the carpet cloak, measurements
were carried out over a wide range of frequencies. The cloaking
behavior was confirmed in our measurements from 10 GHz to 13 GHz, as
shown in Figs. 2(d), (e) and (f). The cloak is excepted to work very
well within a more wide frequency band although it cannot be
verified experimentally due to limitations of the measurement
apparatus. Figs.2(d)-2(f) show similar cloaking behavior to the map
taken at 10 GHz in Fig. 2(c).


The compact-sized carpet cloaks discussed above only work in the
presence of ground-plane. Next, we consider a compact-sized and
broadband free-space cloak based on the electromagnetic mirror
principle although the free-space cloak works only for a specified
direction, as suggested in [4, 7]. Such a kind of directive
free-space cloaks can also find a lot of specific applications like
air flight, space exploration, etc. For practical reason, here we
choose the outer boundary of the free-space cloak as a prolate
cylinder, whose long axis is 125 mm and short axis is 60 mm (see
Fig. 3). Starting from the compact-sized carpet cloak design shown
in Fig. 3(a), which is similar to Fig. 1(a), the ground plane is
removed and the cloaked region in company with the object is
mirrored on $y=0$. As a result, a metallic diamond-shaped object is
surrounded with a free-space compact-sized cloaking material, as
shown in Fig 3(b). Figs. 3(c) and 3(d) illustrate the photos of the
free-space cloak and the diamond-shaped conducting object in
free-space background. In the experiment, a foam, whose
electromagnetic properties are similar to free space, was used to
fix the metamaterials. Since the refractive indices on two sides of
the cloak along the $x$ axis are almost one as marked by the red
dashed lines in Fig. 3(b), such areas can be designed as free space.
The sample is fabricated by using non-resonant metmaterial unit
elements similar to the compact-sized carpet cloak presented
earlier.

To visualize the performance of the compact-sized and broadband
free-space cloak, we make full-wave simulations based on the
finite-element method. In principle, an incident electromagnetic
wave would perceive a perfectly cloaked object as a metallic surface
in the $y=0$ plane with infinitesimal thickness. Such a device will
scatter the impinging wave significantly for most incident angles,
except when the wave is propagating along the $x$ direction. The
simulation results are illustrated in Figs. 4(a) and (b) when a
plane wave is impinging upon a conducting diamond-shaped object
without and with the free-space cloak. A very strong shadow is
observed behind the bare object, along with a weak backward
scattering pattern due to the sharp edge of the diamond object (Fig.
4(a)). Subsequently, the designed cloak is placed around the diamond
object. We observe that there is no shadow left as the wavefronts
are bending and recomposing on the back of the object, with only a
slight distortion, as shown in Fig. 4(b). Hence, the object would be
effectively cloaked.

The measurement results of the diamond-shaped metallic object
without and with the free-space directive cloak are shown in Figs.
4(c) and 4(d) when the collimated beam is incident horizonally at 10
GHz. Obviously, the measured field patterns have excellent
agreements to the simulation results. From Fig. 4(d), we observe
that when the wave reaches the left corner of the diamond-shaped
metallic object, the index gradient is increasing away from the
boundary of the object, and hence the wave bends away from the
object. Then, near the tip of the object, the wave bends back
towards the object because the index gradient is reversed. Finally,
due to the increasing index gradient away from the object near the
right corner, the wave is bent away from the object, and then return
back to their original propagation direction.

To verify the broadband properties of the free-space cloak,
measurements were implemented over a wide range of frequencies. The
cloaking behavior has been confirmed in these measurements from 8 to
12 GHz. We illustrate the broad bandwidth of the cloak with the
field maps taken at 8 GHz in Fig. 4(e), 9 GHz in Fig. 4(f), 11 GHz
in Fig. 4(g) and 12 GHz in Fig. 4(h), which show similar cloaking
behavior to the map taken at 10 GHz in Fig. 4(d). This method would
be increasingly favorable for compact-sized cloaking objects in
free-space background when considering its design simplicity, since
only a few blocks of all-dielectric materials are required in order
to achieve broadband and low-loss performance.


Both compact-sized carpet cloak and directive free-space cloak are
designed based on the isotropic and non-resonant unit elements,
hence the presented cloaks are broadband and low-loss. The agreement
between the measured field patterns and the theory [7] provides
convincing evidence that metamaterials can indeed be widely applied
in practice. The experimental demonstration of compact-sized
invisibility cloaks represents a major step towards real
applications of invisibility devices.

This work was supported in part by the National Science Foundation
of China under Grant Nos. 60990320, 60990324, 60671015, 60871016,
and 60901011, in part by the National Basic Research Program (973)
of China under Grant No. 2004CB719802, in part by the Natural
Science Foundation of Jiangsu Province under Grant No. BK2008031,
and in part by the 111 Project under Grant No. 111-2-05.

\newpage

\section*{{{\bf List of Figure Captions}}}

\noindent \textbf{Fig. 1:} {The design for a compact-sized carpet
cloak in free-space background. (a) Metamaterial refractive index
distribution of the complete carpet cloaking region in which the
mesh lines indicate the quasi-conformal mapping. The compact-sized
cloaking region is shown within the box. (b) Expanded view of the
compact-sized cloaking region in which the refractive indices below
one are all designed as 1. (c) Photograph of the fabricated
metamaterial carpet cloak and the concealed car model. (d) The
design of non-resonant elements and the relation between the unit
cell geometry and the effective index. The dimensions of the
metamaterial unit cells are $a=3$ mm, $w=0.2$ mm, $l=h+2w$ mm, and
$h$ varying from 0 to 2.2 mm.}

\vskip 5mm

\noindent \textbf{Fig. 2:} {(color online) Measured electric-field
mapping of (a) the ground plane, (b) triangular metallic bump, and
(c) ground-plane cloaked bump when collimated beam is incident at
10GHz. The rays display the wave propagation direction, and the
dashed line indicates the normal of the ground in free space and
that of the ground-plane cloak in the transformed space. (d)
Collimated beam incident on the ground-plane cloaked bump at 11 GHz.
(e) Collimated beam incident on the ground-plane cloaked bump at 12
GHz. (f) Collimated beam incident on the ground-plane cloaked bump
at 13 GHz. Far-field patterns when collimated beam is incident at 10
GHz on (g) the ground plane, (h) the triangular metallic bump, (i)
the carpet cloaked bump.}

\vskip 5mm

\noindent \textbf{Fig. 3:} {(color online) The transformation
optical design for the directive free-space cloak. The metamaterial
cloak region is embedded in free-space background. (a) Refractive
index distribution of half free-space cloak in which the mesh lines
indicate the quasi-conformal mapping. (b) Expanded view of the
free-space cloaking region in which the refractive indices below one
are all designed as 1. (c) The photo of the fabricated metamaterial
sample. (d) The side elevation of the fabricated metamaterial
sample.}

\vskip 5mm

\noindent \textbf{Fig. 4:} {(color online) Simulated electric-field
distribution of the directive free-space cloak when a plane wave is
incident at 10 GHz on (a) the bare metallic diamond-shaped object,
(b) on the cloaked diamond-shaped object at 10 GHz. The rays display
the wave propagation direction. Measured electric-field mapping of
the directive free-space cloak when a plane wave is incident on (c)
the bare metallic diamond-shaped object at 10 GHz, (d) the cloaked
diamond-shaped object at 10 GHz, (e) the cloaked diamond-shaped
object at 8 GHz, (f) the cloaked diamond-shaped object at 9 GHz, (g)
the cloaked diamond-shaped object at 11 GHz, (h) the cloaked
diamond-shaped object at 12 GHz.}

\newpage

\begin{figure}[t]
\centerline{\includegraphics[width=11.2cm]{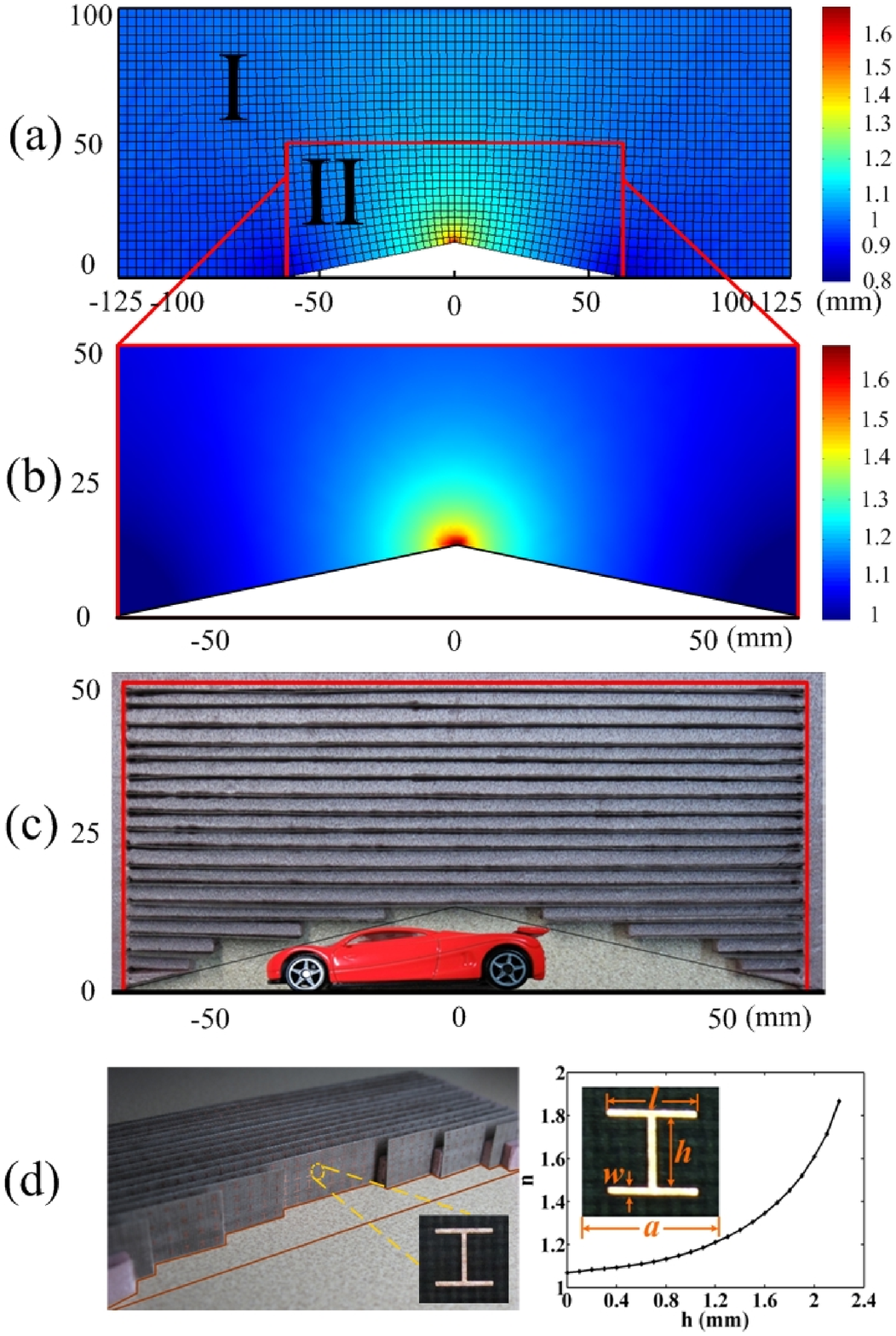}}
\caption{}\label{fig1}
\end{figure}

\begin{figure}[t]
\centerline{\includegraphics[width=15cm]{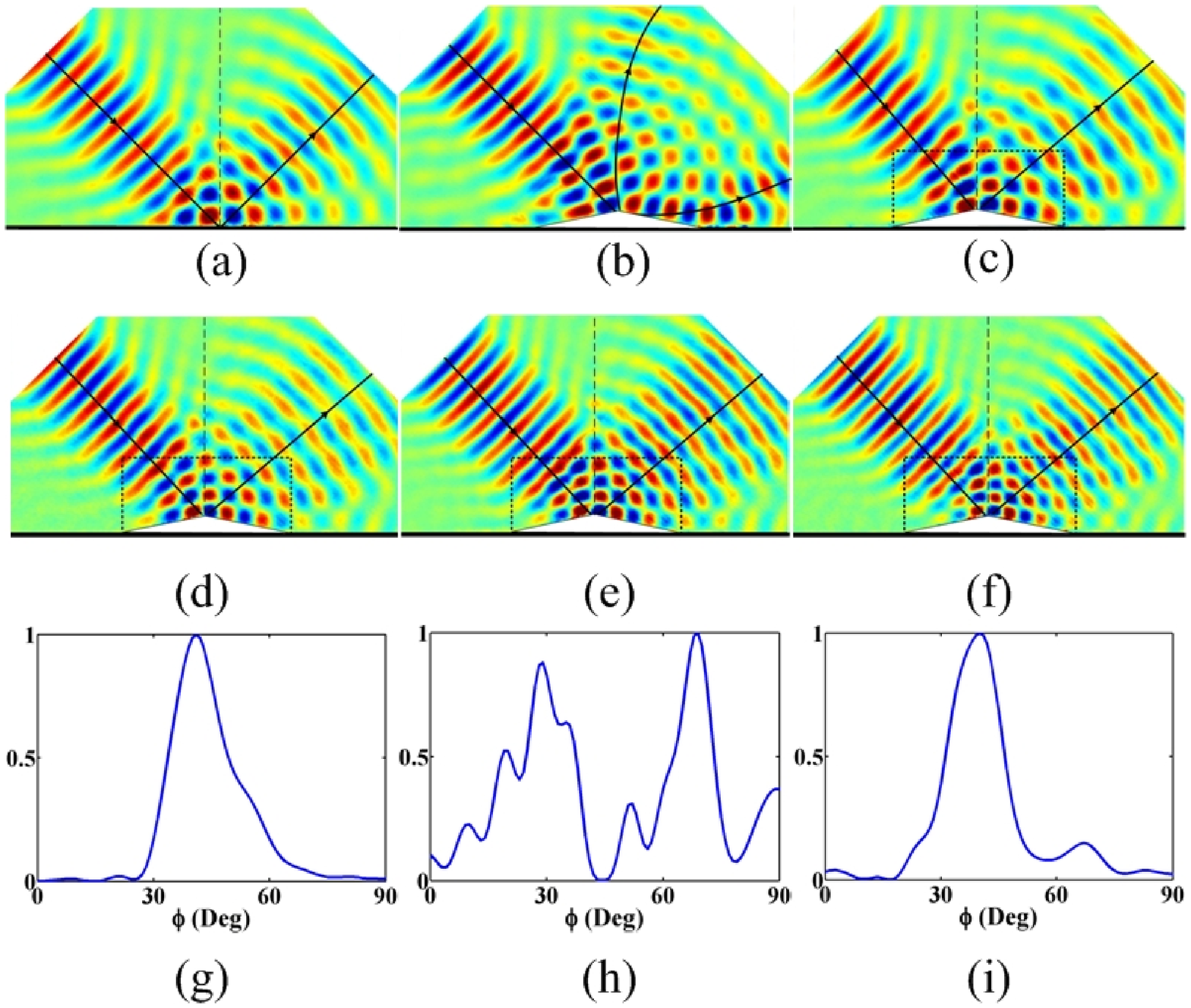}}
\caption{}\label{fig2}
\end{figure}

\begin{figure}[h,t,b]
\centerline{\includegraphics[width=11.5cm]{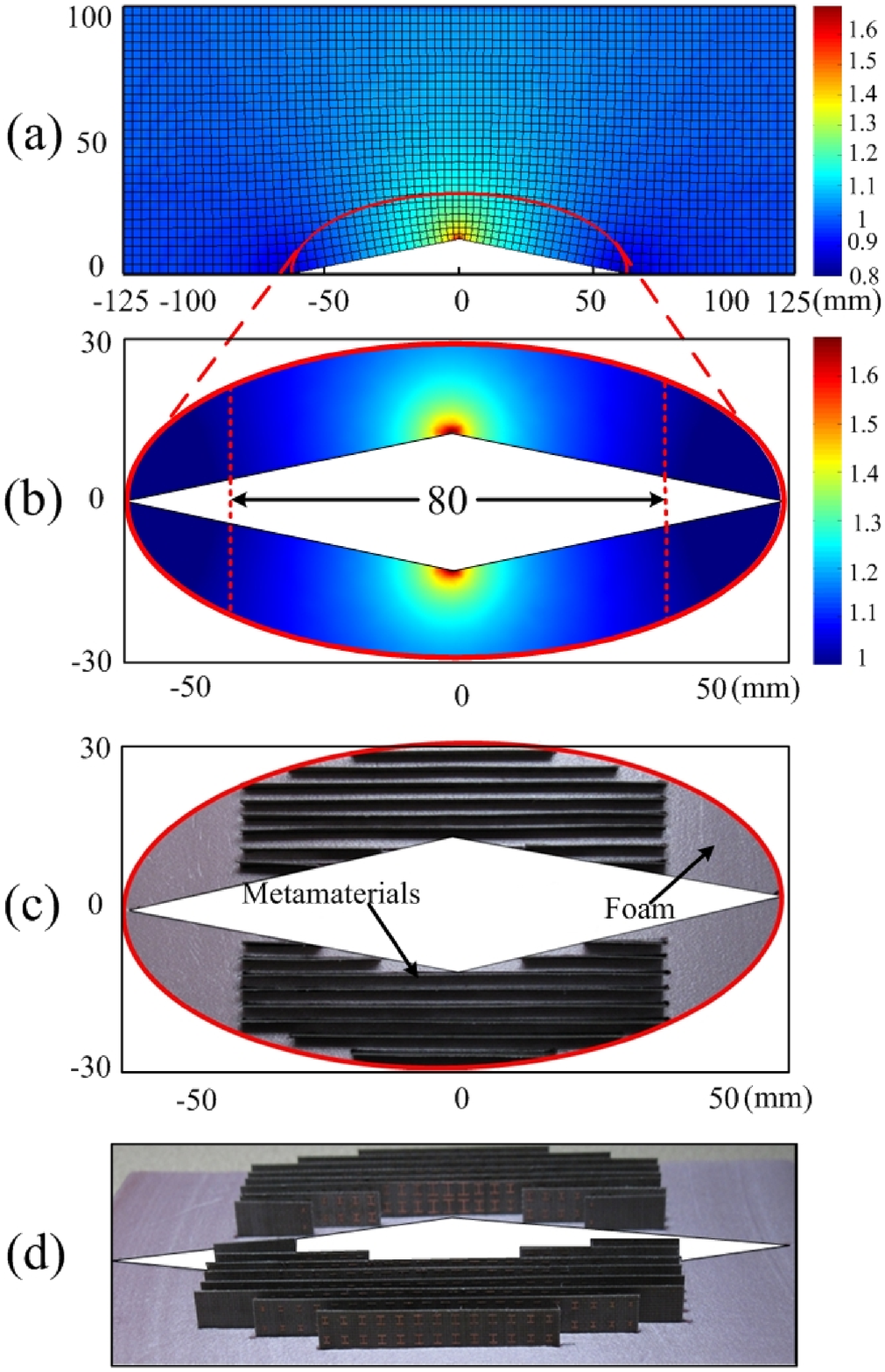}}
\caption{}\label{fig3}
\end{figure}

\begin{figure}[h,t,b]
\centerline{\includegraphics[width=15cm]{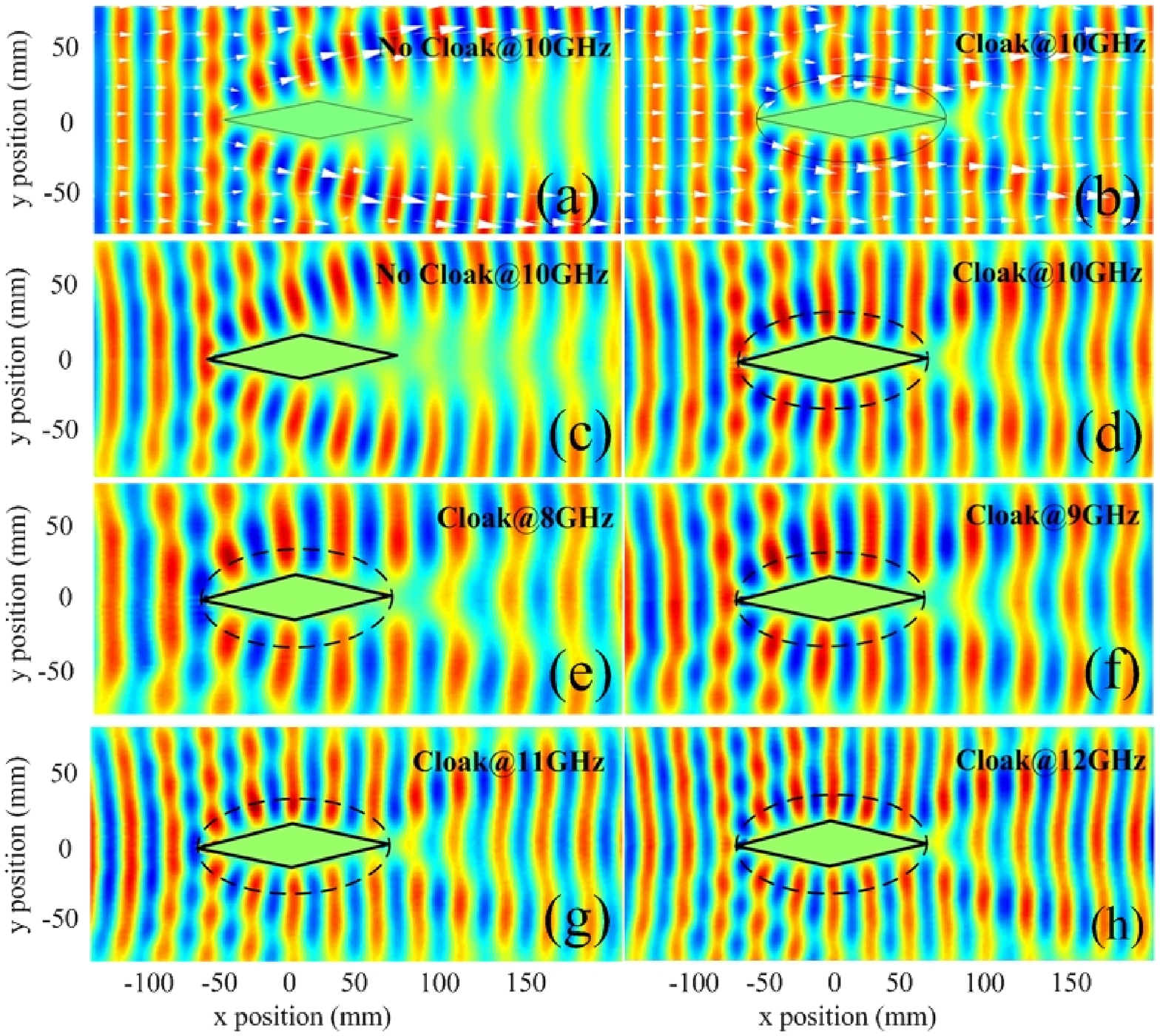}}
\caption{}\label{fig4}
\end{figure}

\end{document}